\documentclass[]{emulateapj}
\usepackage{apjfonts}
\usepackage{graphics,amssymb}
\usepackage{bm}

\shortauthors{Woodring et al.}
\shorttitle{Analyzing and Visualizing Cosmological Simulations with ParaView}

\begin{document}
\journalinfo{The Astrophysical Journal}
\submitted{Submitted to the Astrophysical Journal Supplements}


\vbox to 0pt{\vss
\hbox to 0pt{\hskip 440pt\rm LA-UR-10-06301\hss}
\vskip 25pt}

\title{Analyzing and Visualizing Cosmological Simulations with ParaView}
\author{Jonathan~Woodring\altaffilmark{1}, Katrin~Heitmann\altaffilmark{2},
        James Ahrens\altaffilmark{1}, 
        Patricia Fasel\altaffilmark{3}, Chung-Hsing Hsu\altaffilmark{4},
        Salman~Habib\altaffilmark{5},
        and Adrian~Pope\altaffilmark{2,5}}

\affil{$^1$ CCS-7, CCS Division, Los Alamos National Laboratory, Los
Alamos, NM 87545}
\affil{$^2$ ISR-1, ISR Division,  Los
Alamos National Laboratory, Los Alamos, NM 87545}
\affil{$^3$ CCS-3, CCS Division, Los Alamos National Laboratory, Los
Alamos, NM 87545}
\affil{$^4$ Oak Ridge National Laboratory, Oak Ridge, TN 37831}
\affil{$^5$ T-2, Theoretical Division, Los
Alamos National Laboratory, Los Alamos, NM 87545}

\begin{abstract} 
  The advent of large cosmological sky surveys -- ushering in the era
  of precision cosmology -- has been accompanied by ever larger
  cosmological simulations. The analysis of these simulations, which
  currently encompass tens of billions of particles and up to trillion
  particles in the near future, is often as daunting as carrying out the
  simulations in the first place. Therefore, the development of very
  efficient analysis tools combining qualitative and quantitative
  capabilities is a matter of some urgency. In this paper we introduce
  new analysis features implemented within ParaView, a parallel,
  open-source visualization toolkit, to analyze large $N$-body
  simulations. The new features include particle readers and a very
  efficient halo finder which identifies friends-of-friends halos and
  determines common halo properties. In combination with many other
  functionalities already existing within ParaView, such as histogram
  routines or interfaces to Python, this enhanced version enables
  fast, interactive, and convenient analyses of large cosmological
  simulations. In addition, development paths are available for future
  extensions.

\end{abstract}


\section{Introduction}

During the last two decades, cosmological measurements and predictions
have advanced from the level of estimates to precision determinations
-- at better than the 10\% level -- of the major cosmological
parameters. In the next decade, ongoing and upcoming surveys such as
the Sloan Digital Sky Survey III, PanStarrs, the Dark Energy Survey,
the Large Synoptic Survey Telescope, the Joint Dark Energy Mission, or
Euclid, to name a few, promise improvements in the measurement state
of the art by yet another order of magnitude. At the same time,
theoretical predictions at the same or better levels of accuracy are
needed to fully exploit the information available from these surveys.
Predictions of this quality can only be obtained by high-performance
simulations which cover cosmological volumes representative of those
observed by the surveys. At the same time, the simulations must possess
high enough mass and force resolution to reliably resolve dark matter
halos which host galaxies. For gigaparsec cubed volumes, the
requirements translate to tens to hundreds of billions of simulation
particles.

\begin{figure}[b]
\centerline{
\includegraphics[width=2.9in,angle=90]{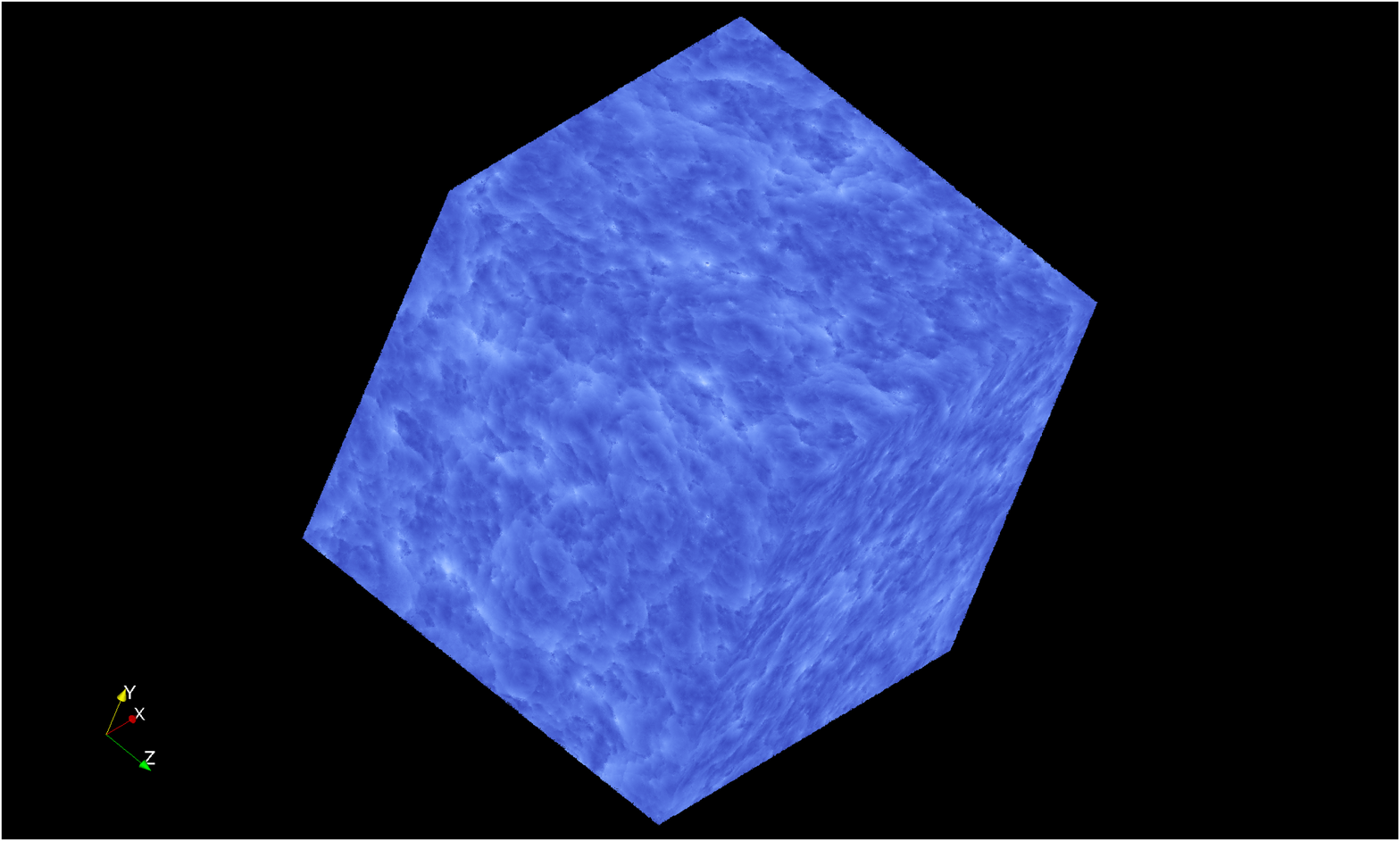}
\includegraphics[width=2.9in,angle=90]{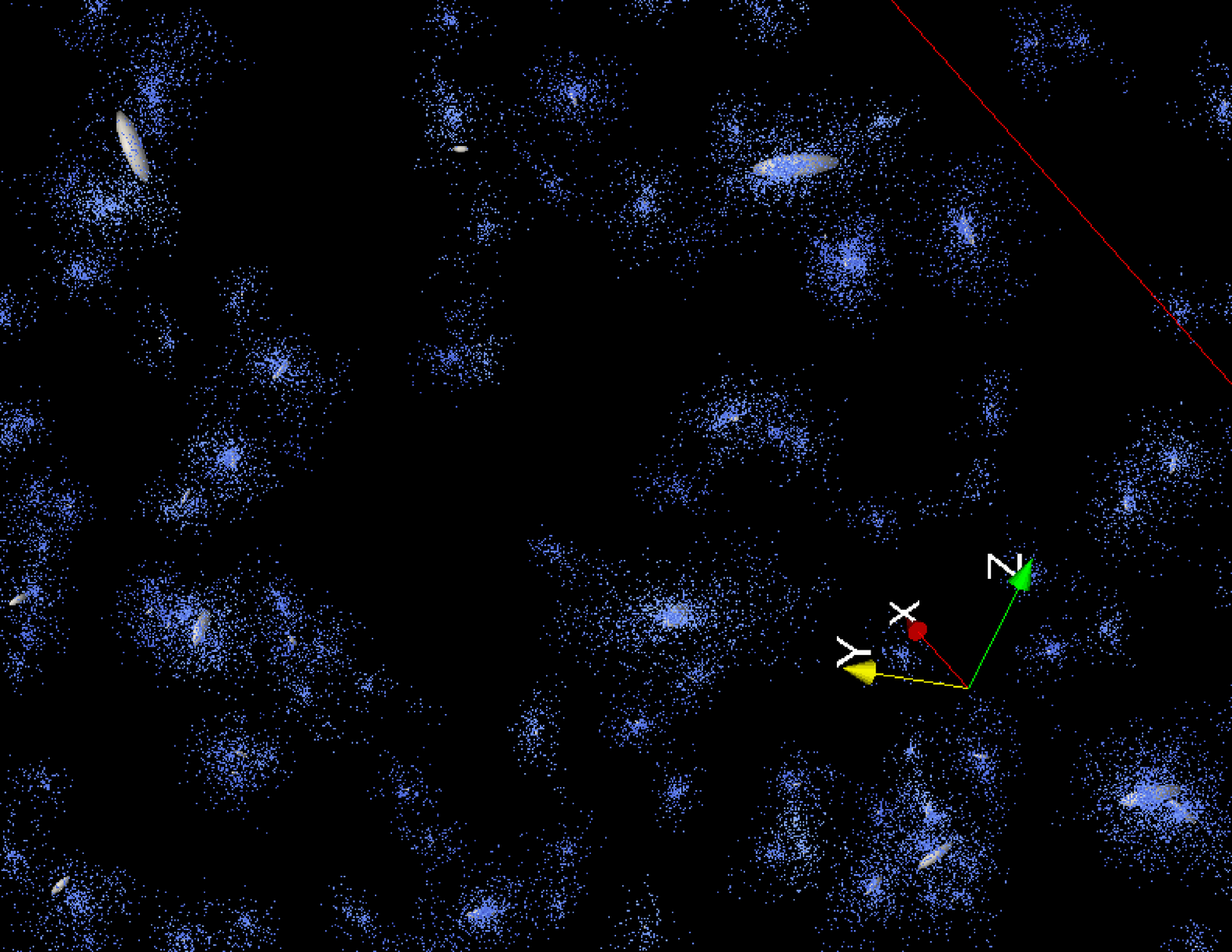}}
\caption{\label{billion}ParaView visualization of a billion particle
  simulation from the Coyote Universe suite~\citep{CoyoteIII}. Left:
  visualization of all particles colored by their velocities; right:
  visualization of a sub-volume showing particles in halos and the
  halo centers.} 
\end{figure} 

In the past several years, this simulation challenge has been attacked
from different perspectives. A few `hero' simulations have been
carried out: The Millennium simulation from 2005 with $\sim$10 billion
particles and the two ``Horizon'' simulations from 2009 with $\sim$70
billion particles each~\citep{horizon1,horizon2}, but with lower force
resolution, are prominent examples. Another approach is to run more
moderately sized simulations (still one to ten billion particles each)
but with many realizations and different volumes for one cosmological
model. This allows for efficient gathering of statistics and the
determination of covariances (e.g. the LasDamas Project by
\cite{mcbride2010} and the MICE simulations by \citealt{MICE}). Such
simulations can also be carried out for different cosmologies
(e.g. the Coyote Universe by \citealt{CoyoteI,CoyoteII,CoyoteIII}) to
explore the cosmological parameter space and derive predictions for
different statistics of interest. Both approaches generate a large
amount of data and the analysis task is often as demanding as carrying
out the simulations themselves. It is therefore very desirable to
develop efficient, flexible, versatile, and easily extendable tools to
help with the analysis task. Taking this thought one step further, an
analysis tool that combines quantitative and qualitative features
would be particularly convenient; such a tool should allow
visualization and analysis of the data at the same time, and allow
user-customizable features and extensions.

\begin{figure*}[t]
\centerline{
\includegraphics[width=6in]{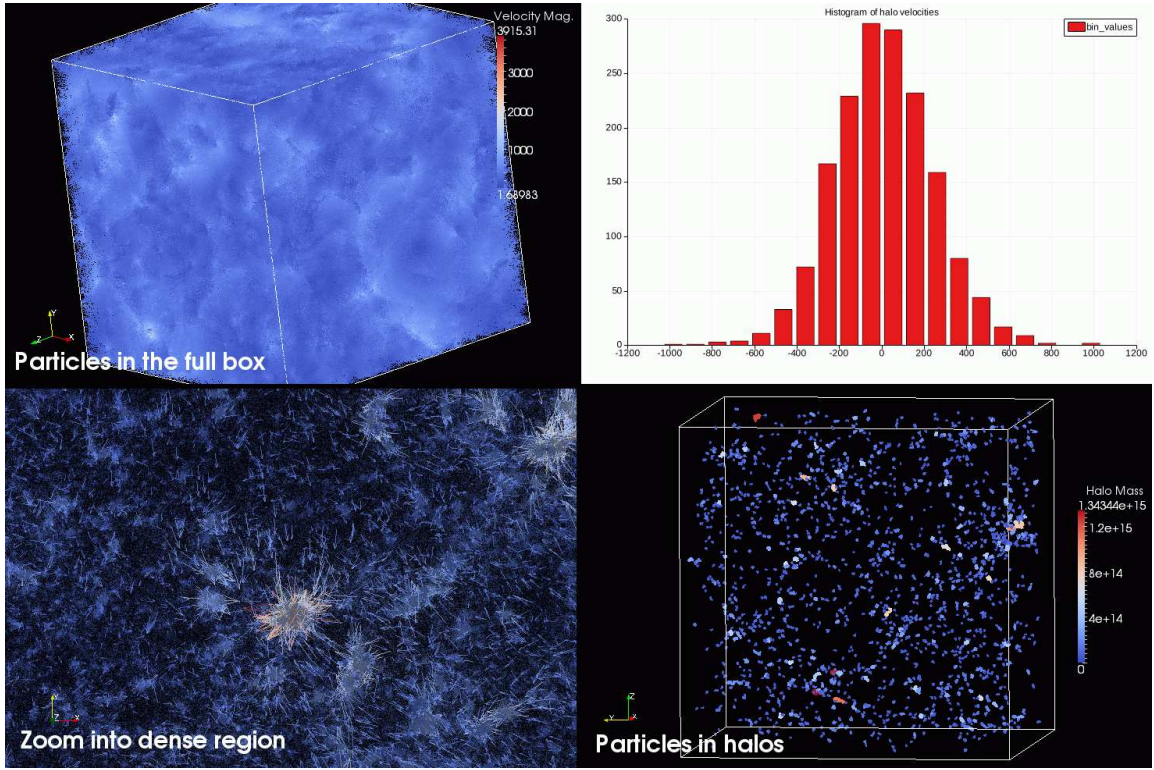}}
\vspace{0.3cm}
\caption{\label{screenshot}Examples of different analysis capabilities
  of ParaView. The simulation shown here is described in Section
  \ref{sims}.  The upper left panel shows all the particles in the
  box colored with respect to their velocities, the lower left panel
  shows a zoom-in into a dense region. The upper right panel shows a
  histogram of halo velocities. The lower right panel shows only the
  particles that reside within halos.}
\end{figure*}

Over the last several years we have developed a visualization and
analysis tool based on ParaView, an open-source, parallel
visualization framework. Figure~\ref{billion} shows an example
visualization of a billion particles carried out on 128 processors. We
used ParaView for some of the analyses carried out in
\cite{CodeCompare}. In \cite{Vis2010} we demonstrated the tool's
efficiency for scientific investigations by analyzing the formation of
halos over time. In the current paper, we introduce the new features
we have implemented into ParaView to analyze and visualize large
cosmological $N$-body simulations. These features are included in the
latest ParaView~3.8
release\footnote{http://www.paraview.org/}. ParaView is a very
convenient platform for several reasons: (i) it
is parallel and therefore well suited for very large data sets, (ii)
some general data analysis routines already exist within it, (iii) it is open
source, and (iv) it can be coupled with relative ease to existing codes
and includes interfaces to programming languages, e.g., Python.

We have implemented two different particle readers into ParaView, one
based on the `cosmo' format introduced in \cite{heitmann05}, the other
following the `{\small GADGET}'
format\footnote{http://www.mpa-garching.mpe.de/{\small
    GADGET}/users-guide.pdf} widely used by the cosmology community
and the native data format of the cosmology code {\small GADGET-II}
\citep{springel}. The cosmo format is a simple binary format for
storing particle positions, velocities, tags, and mass. The {\small
  GADGET} format allows for more flexibility and information, although
currently the ParaView reader only reads the $N$-body particles from
the {\small GADGET} files. A future ParaView release will include the
option of reading gas information as well. In addition to the readers,
we have implemented a very efficient parallel halo finder that
supports the friends-of-friends (FOF) algorithm. Combined with the
analysis features already available, such as histogram routines,
density routines, comparative visualization, movie options and so on,
ParaView provides a convenient and flexible visualization and analysis
environment.

The paper is organized as follows. First, we give a short overview of
ParaView and briefly describe the new cosmology modules that have been
implemented. In Section~\ref{fof} we provide a more detailed
description of the parallel halo finder and the halo properties
available. We give an example for how ParaView can be used to analyze
cosmological simulations in Section~\ref{example} and conclude in
Section~\ref{conclusion}. We give a brief introduction on the ParaView
GUI and the use of the cosmology filter in the Appendix.

\section{ParaView}

ParaView \citep{paraview} is a general-purpose, open-source,
scientific visualization server technology built on top of VTK (the
Visualization Toolkit) \citep{vtk}. Through the ParaView server,
visualization of large-scale data is possible by parallelism and data
streaming on a scalable server backend. The default ParaView
application runs a built-in single-threaded ParaView server for small
visualization tasks. For larger data, it can connect to a remote
Message Passing Interface (MPI) parallel ParaView server backend that
is running on a visualization cluster or parallel supercomputer.

The process of visualization within ParaView consists of constructing
VTK visualization pipelines of readers and filters, where outputs are
implicitly connected to render views. Readers allow data to be
imported into the pipeline, while filters allow data to be analyzed
and manipulated through processing. Render views provide visual
representations of data that a user can interact with. Visualization
pipelines are constructed through a front-end interface, which is
capable of communicating with the ParaView server, and are executed on
the data by the server. Visual results are displayed by the front-end
from images or geometry sent back by the ParaView server after
processing data using a constructed visualization pipeline.

A variety of tools and languages can interface with the ParaView
server backend to analyze and visualize data.  The images we show in
the following were created using the default ParaView graphical user
interface client available from the ParaView website. The default
ParaView GUI is a Qt application, with Python scripting support, that
is available on Windows, OS X, Linux, and on any other platform that
is able to compile C++ code with the Qt framework \citep{Qt}. ParaView
is also capable of performing visualization and analysis through other
front-ends, such as task specific visualization tools built on top of
the ParaView server language bindings in Python, Tcl/Tk, C++, Java,
and Javascript. The ParaView parallel server backend compiles on any
platform capable of compiling C++ and MPI code. Information on using
ParaView and downloading source and binaries is located at {\em
paraview.org}.

\subsection{Parallel Reading}
\label{reading}

Assuming that the ParaView server is running in parallel (the reader
will works in serial mode as well), the first task to visualize
cosmology data is to correctly read the data and distribute it among
the processors. The ParaView reader we have implemented for the
``cosmo'' format has also been extended to work with the ``{\small
  GADGET}'' format (for $N$-body particles). Particle files can be
single monolithic files or multiple files generated per-processor
during the simulation.

The first task the reader performs is assignment of processors to
regions in space, such that a processor will be assigned a contiguous
block in space. We use a three dimensional spatial decomposition. Each
processor will eventually obtain all of the particles in that
space. The second task is reading the particle file or files.  If
there is a single file, all of the processors will read into memory a
linear portion of the particle list, temporarily taking ownership of
the particles in that segment of list. Likewise, if there are per
process files, each processor will read a file assigned to it, and
take temporary ownership of the particles located in that file.  If
there are fewer or more files per process, the files are divided such
that each processor reads an equal number of particles.

Next, processors take ownership of the actual particles that belong to
them through communication of particles, i.e., moving the particles to
the correct processor that owns the spatial region containing the
particle. In the first step, the processors examine the particles
currently in memory and separate out the particles that belong on that
processor, from the particles that do not belong. Next, all of the
processors simultaneously send a buffer of particles that do not
belong to the next rank processor, while receiving a buffer from the
previous rank processor. Each processor will examine the buffer
received and take out the particles that belong to it. If the number
of files is greater or equal the number of processors, this process is
repeated for $p - 1$ rounds, where there are $p$ processors, until all
of the particles are in memory on the correct processor. In the case
of more processors than files, the round robin circles are smaller so
that every processor reads a file if possible. For example, for 16
files and 64 processors, 4 processors will read each file and the
round robin chain is $[(p / 4) - 1]$ in size.

Finally, in order to perform correct halo finding per-process, we
allow for spatial overlap in the per-process volumes and duplicate the
particles across overlap regions. Given that the entire space is
divided into blocks and there is wraparound on boundaries (because of
periodic boundary conditions), each processor will have 26 neighbor
processors. With an appropriate overlap boundary size, as explained in
Section~\ref{fof}, each processor can determine the volume overlap or
intersection regions with its neighbors. The duplicate or ``ghost''
particles in the overlap regions are communicated to each of spatially
contiguous neighbor processes to expand the volume of each process.

\subsection{Parallel Filtering and Rendering}

The data, as it is read in, is treated as a parallel VTK unstructured
point data set in the visualization pipeline. The ``cosmo'' format
provides point position, velocity, mass, and a tag available as data
attributes (fields, or variables) per point in a binary file. The data
at this point can be rendered as is, using parallel point rendering
colored by data attribute, or it can be analyzed and manipulated
through various parallel VTK filter modules before rendering.

There are many built-in VTK filters available in ParaView, and it
takes a small amount of effort by an expert user to expose an existing
VTK filter in ParaView that is not already available by default.  It
involves creating an XML plugin to tell ParaView how to interface with
the VTK internals. Some useful filters available by default in
ParaView are the Calculator, Threshold, and Glyph filters. The
Calculator filter allows new derived fields to be calculated on the
fly from existing scalar, vector, and tensor fields using a
mathematical expression. The Threshold filter discards data points
that do not lie within a given range for a data value. The Glyph
filter generates new geometry per point that can be scaled and
oriented by attributes, such as spheres whose size is dependent on
mass or arrows that point in the direction of the velocity and scaled
by the velocity magnitude. Examples are shown in
Section~\ref{example}. Finally, if a desired filter does not already
exist in VTK, ParaView includes the capability to script new filters
in Python.

\subsubsection{An Efficient, Parallel, Friends-of-Friends Halo Finder}
\label{fof}

An important component of the new ParaView cosmology capabilities is a
very efficient parallel halo finding filter. The base implementation
is a fast serial FOF halo finder, with parallel integration added. For
finding FOF halos, the user can specify the linking length and the
minimum number of particles defining a halo. ParaView returns a halo
catalog containing halos with average position, average velocity,
one-dimensional velocity dispersion, and mass for each halo.
Optionally, the original particle list can be also be annotated with
the halo information that each particle belongs to.

In order to achieve performance goals for the halo finding algorithm,
we first developed a new serial halo finder~\citep{Vis2010}. A naive
implementation of an FOF finder would check each and every particle
pair; given $n$ particles, therefore requiring ${\cal O} (n^2)$
operations, clearly an unacceptable scaling. To reduce the number of
operations, we use a balanced kd-tree. A balanced kd-tree is a binary
space partitioning data structure that organizes points in a
$k$-dimensional space in such a way that the number of points in a
subtree at each level are equal. Building a fully balanced tree from
$n$ points takes ${\cal O} (n\log n)$ operations.
 
\begin{figure}[t] \centerline{
    \includegraphics[width=3.2in]{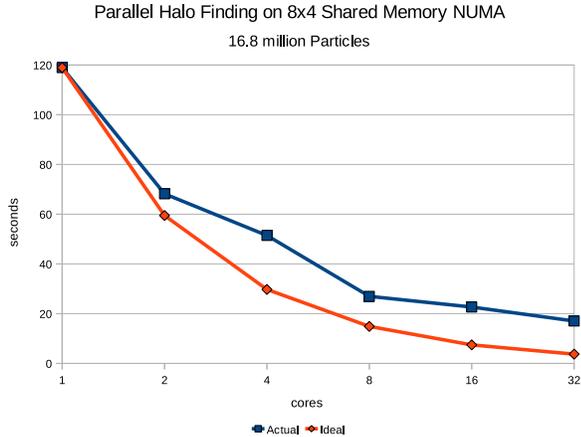}} \vspace{0.3cm}
  \caption{\label{timing}Strong scaling of the parallel halo finder in
    ParaView. The red curve shows ideal scaling, while the blue curve
    shows the actual timing. The problem size is fixed (256$^3$
    particles) as the number of cores is increased from 1 to 32.
    Scaling is close to ideal; reasons for the departures are discussed
    in the text.} \end{figure}
 
A recursive FOF algorithm starts at the leaf nodes (single particles)
and merges nodes into halos by checking if the particles are within a
given range of each other. As particles are merged into halos,
particle tests are reduced by using subtree bounding boxes as proxies
for points. If a subtree bounding box is too distant, all of the
points can be skipped. Vice versa, if an entire bounding box is close
enough all of the points can be added to a halo.

\begin{figure*}[t]
\centerline{
\includegraphics[width=7.1in]{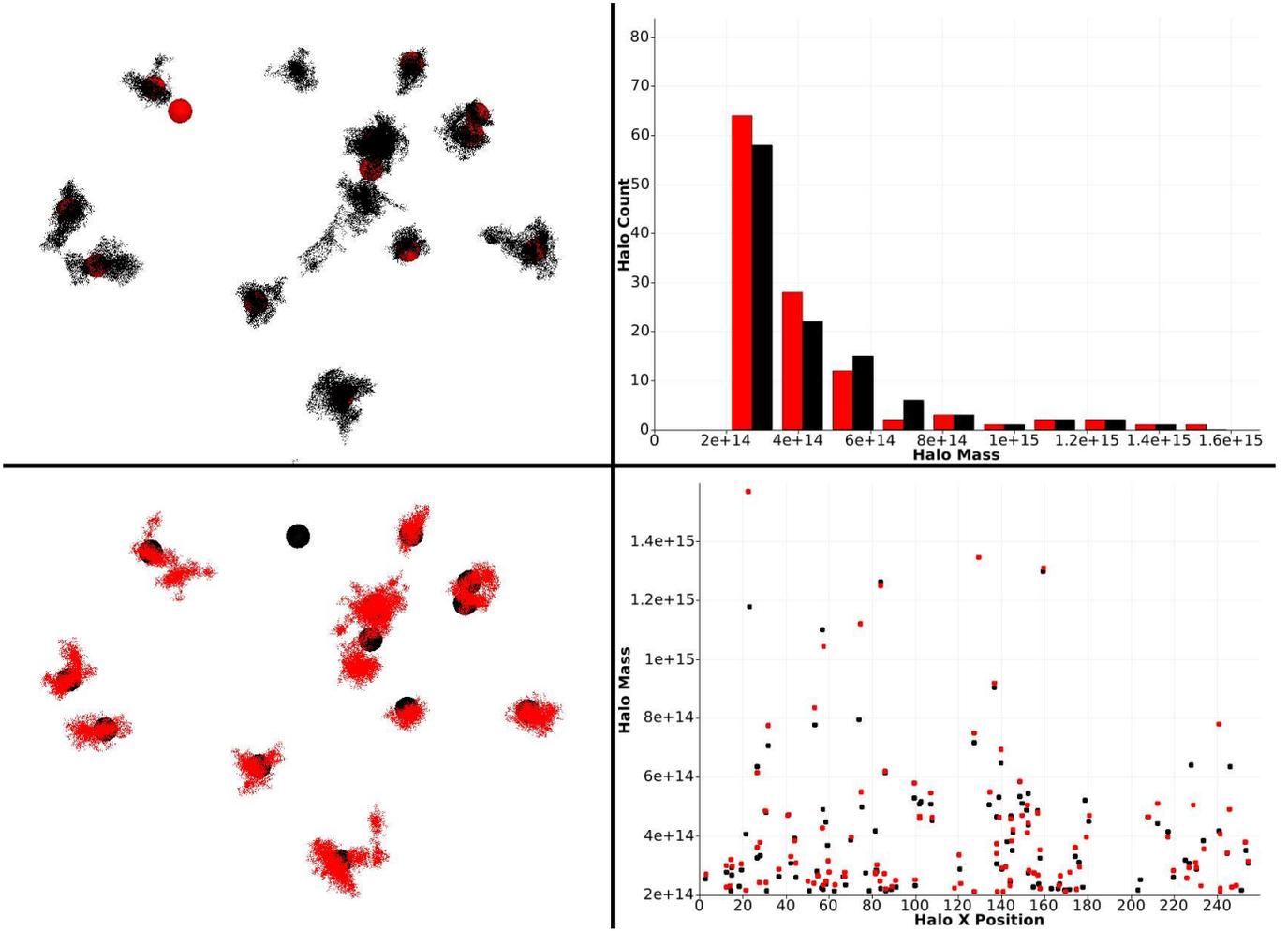}}
\vspace{0.3cm}
\caption{\label{b02} Force resolution and halo formation test: All
  results shown are for a linking length $b=0.2$ and more than 3000
  particles per halo. Results from the high resolution run are shown
  in red, results from the low resolution run are shown in black. Left
  upper panel: particles in halos from the low resolution run (black)
  and halo centers of halos from the low resolution run (red spheres)
  in a sub-volume of the simulation. The left lower panel has the same
  information but with particles from the high resolution run (red)
  and centers (black spheres) from the low resolution run. The
  overlinking problem can be seen for several halos: a very obvious
  example is a halo in the central region that links several
  structures together in the low resolution simulation. Also, the
  halos that are absent in the low resolution simulation are clearly
  at the lower mass end, the red centers in the upper left plot
  correspond to small halos that can be identified in the lower left
  plot. Note though, that these halos are not {\em missing} in the low
  resolution run, but rather not showing up in the plot because of our
  mass threshold of 3000 particles -- in other words they are there
  but are below our mass threshold. Several points are apparent from
  this comparison: the low resolution halos are less concentrated,
  halos in the lower resolution run are more often overlinked, and
  some of the halos found in the higher resolution run are missing in
  the lower resolution run because they fell below the 3000 particle
  cut. The right upper panel shows a histogram of the halo counts as a
  function of mass. For the lowest two mass bins, the high resolution
  run has more halos than the low resolution run.  The lower right
  panel shows the $x$-position versus halo mass, presenting the
  information in the two left panels in condensed, but more
  quantitative form. It also shows that the positions of the halos in
  both simulations are in reasonable agreement.}
\end{figure*}

Once a fast serial FOF finder has been built, the next step is to
implement efficient parallelization. We use a straightforward strategy
by dividing the simulation volume into per-processor sub-volumes and
allow these sub-volumes to overlap, as described in
Section~\ref{reading}. The overlap length should be larger than the
diameter of the largest halo, usually $\sim 5$~Mpc is a conservative
choice. This is done to ensure that each halo is complete in at least
one processor (overlapped) sub-volume. Next, the algorithm finds all
halos in the sub-volumes. The last step is to ensure that each halo
once is counted only once. When a halo is shared between two
processors across a plane, it is assigned to the processor which has
the halo in the upper plane (this is an arbitrary choice) and is
eliminated from the other. If it is shared by more than one processor,
the information is sent to an arbitration processor that makes the
assignment and informs all other processors.

Strong scaling (execution time as a function of processor
number with fixed problem size) of the halo finder is demonstrated in
Figure~\ref{timing}. The test shown is carried out on a 32 core shared
memory machine with 128 GB of RAM. The test used a 256$^3$ particle
simulation snapshot at $z=0$. The actual timing of the halo finder is
slightly higher than the ideal value, due to several reasons: (i) The
halo finder is not fully load balanced since a large halo would cause
a certain processor to do more work than others. For large volume
simulations this is not a severe problem since not very many
exceptionally large halos form.  (ii) Due to the overlap strategy, the
workload increases as parallelism increases. (iii) The ideal curve
does not account for communication overhead. With these caveats in
mind, the scaling behavior of the halo finder is very good. In
addition, we also carried out a timing test on a $1024^3$ particle
simulation in distributed memory MPI. Results are given in
Table~\ref{tabscal}.

\begin{table}\caption{Halo finder timing for a billion particles}
\begin{center}
\begin{tabular}{c c}
\hline
\# of processors & Time in sec \\
\hline
 64 & 66.6\\
 128 & 32.9\\
 256 & 20.5\\
\hline \hline
\end{tabular}\label{tabscal}
\end{center}
\end{table}

\begin{figure*}[t]
\centerline{
\includegraphics[width=7.1in]{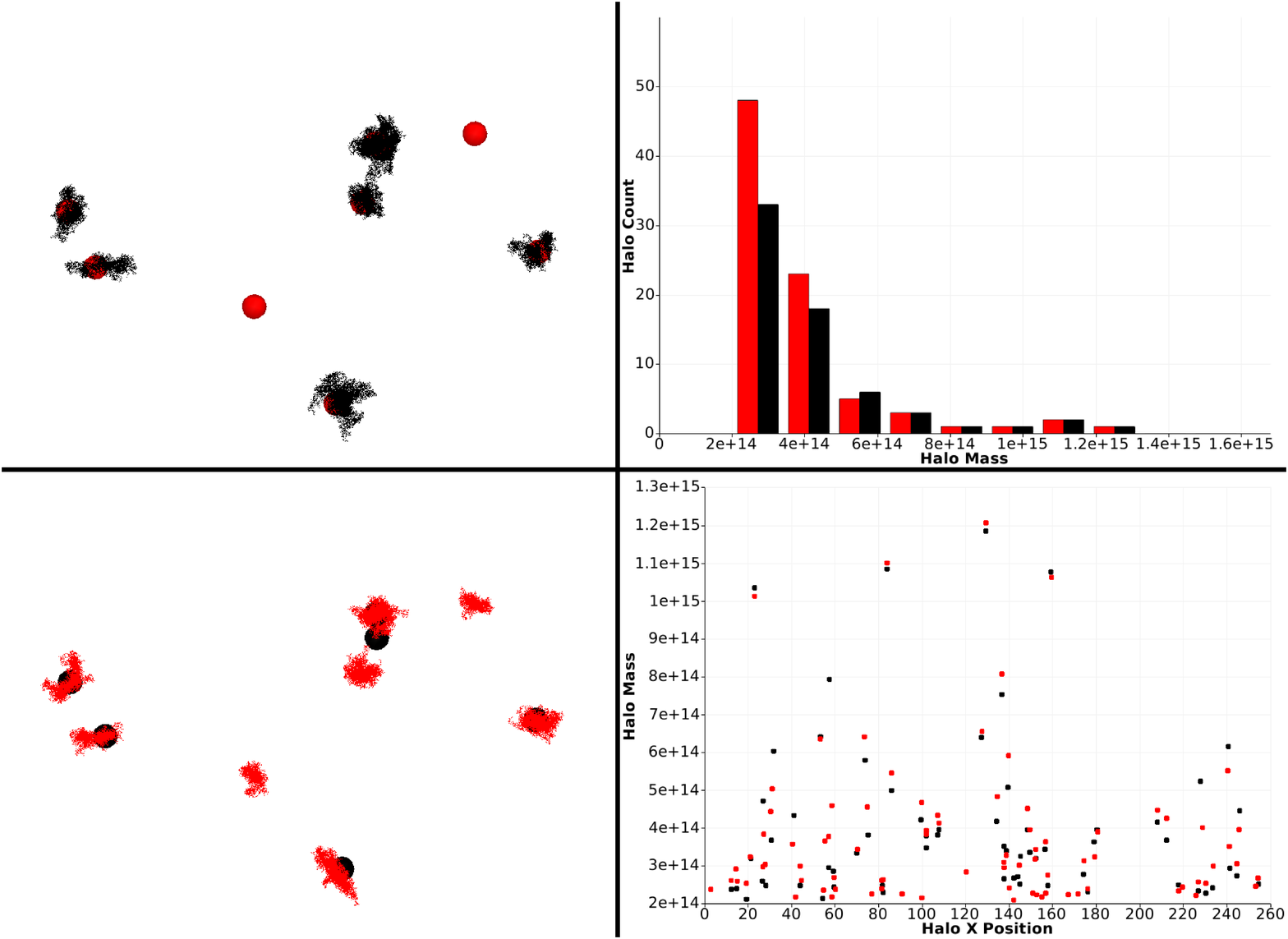}}
\vspace{0.3cm}
\caption{\label{b015}Same as in Figure~\ref{b02} but for a halo-finder
  linking length of $b=0.15$. As expected, the discrepancies in the
  number of halos between the low and high resolution runs further
  worsen (Cf. the halo count vs. mass histogram, top right); the
  ``orphaned'' red spheres in the upper left plot all correspond again
  to smaller halos. But the overlinking problem is essentially
  absent.} \end{figure*}

Halos do not have a uniquely defined notion of `halo center'. Of the
different possibilities, the center of mass is the easiest to 
implement and fastest to run. In this case, the particle-averaged
position of the halo is given by 
\begin{equation}
{\bf x}_{\rm FOF}=\langle {\bf x} \rangle = \frac{1}{n_{\rm
FOF}}\sum_i^{n_{\rm FOF}}{\bf x}_i,
\end{equation}
where $\bf x_i$ is the position of the $i^{th}$ particle in the halo
and $n_{\rm FOF}$ is the number of particles in the FOF halo. The halo
center of mass velocity is determined in an analogous fashion. Because
this is the fastest way of determining the halo center, it is the
default setting we choose for ParaView. Of course this definition has
obvious shortcomings: e.g., if a halo is comprised of two distinct
subhalos, the center of mass will lie in between those subhalos and
not at the center of the more massive subhalo. A more accurate
determination of the halo center is therefore given by either finding
the potential minimum of the halo or to find the particle with the
most neighbors (these two centers are very close and for most halos in
fact identical). In future, ParaView will allow for this option in
addition to determining the center of mass.

\begin{figure}[t]
\centerline{
\includegraphics[width=3.2in]{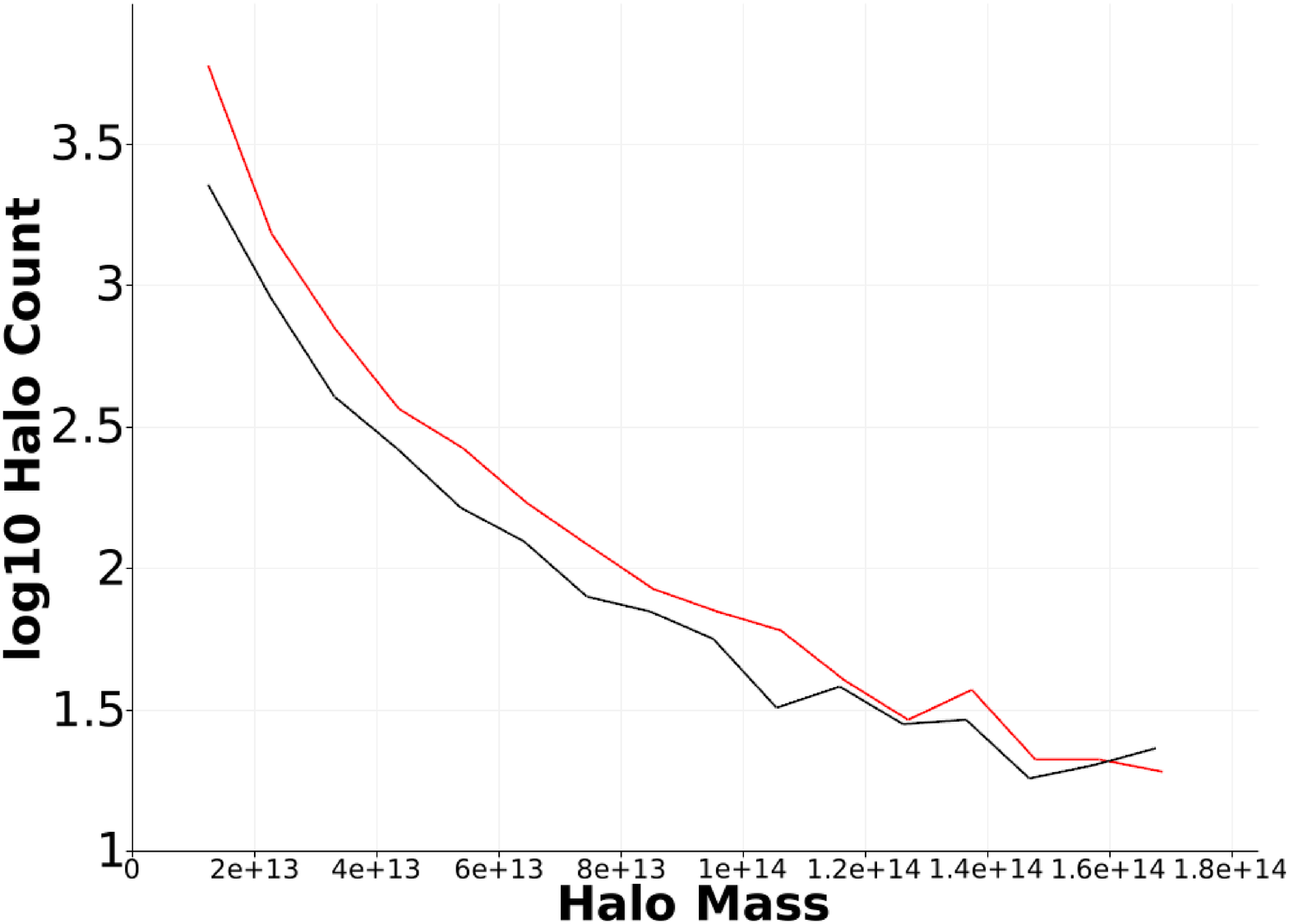}}
\caption{\label{massf}Number count of halos vs. mass for halos with
less than 2500 particles. Black line: low resolution simulation, red
line: high resolution simulation. Following our estimate, the low
resolution simulation has fewer halos than the high resolution
simulation over the whole mass range considered.} \end{figure}

In addition, the halo finder provides a measurement of the
one-dimensional velocity dispersion, given by: 
\begin{equation}
\sigma_{v} = 
\sqrt{
\frac{1} {3}
\left(
\frac {1} {n_{\rm FOF}} \sum_{i}^{n_{\rm FOF}} 
\bf{v}_i \cdot \bf{v}_i
- \bf{v}_{\rm FOF} \cdot \bf{v}_{\rm FOF}
\right)}.
\end{equation}
In future, the set of halo properties calculated by ParaView will be
extended to include e.g. spherical overdensity mass and halo
concentration as well as sub-halo finding. The current 3.8 release is
limited to the aforementioned halo finding features and halo
properties.

\section{Analysis of Cosmological Simulations with ParaView}  
\label{example}

In this section we focus on a simple example to demonstrate how
ParaView can be used to gain better intuition for cosmological
structure formation by {\it visualizing} data sets and at the same
time can be used to {\it analyze} the data sets and to extract
quantitative information. The example we investigate is the effect of
the force resolution in the simulation on the accuracy of halo masses.

The era of precision cosmology poses daunting challenges to
theoretical cosmologists. Accuracy requirements at the 1\% level for
simulations of highly nonlinear processes such as structure formation
demand extremely careful analyses of possible systematic errors in
$N$-body simulations. A powerful probe of cosmology is the mass
function which yields the number of halos as a function of their mass.
The mass function is very sensitive to cosmological parameters and
enables us to, e.g., distinguish between different models of dark
energy. It was pointed out recently by \cite{heidi} and \cite{cunha}
that in order to analyze the data from future cluster surveys we need
predictions for the mass function at the 1\% level of acuracy.
\cite{bhattacharya} find that uncertainties in the measurement of halo
masses at the 2\% level translate into inaccuracies in the mass
function in the cluster mass regime at the 5-10\% level. Therefore,
understanding systematic biases of halo masses due to numerical
shortcomings is a significant issue (aside from problems with the
physical modeling itself) if we aim to predict the mass function at
high accuracy.

Two major sources of numerical inaccuracy in determining halo masses
are limitations in mass and force resolution. As pointed out by
\cite{warren}, and later investigated by \cite{lukic} for idealized
Navarro-Frenk-White (NFW) halos \citep{NFW}, undersampling halos with
particles, i.e. insufficient mass resolution, leads to a systematic
increase in the halo mass. The effect of insufficient force resolution
is twofold: (i) The boundaries of the halo are not as tight, therefore
more particles from the surrounding area will be linked to the halo
and lead to a mass increase. (ii) The concentration of the halo is
considerably lower and less mass resides in the halo center, which can
lead to a decrease in the total mass. We will use ParaView to
investigate these force resolution effects in more detail and show how
the use of ParaView can provide an intuitive understanding as well as
yield quantitative results.

\subsection{The Simulations}
\label{sims}

Our test analysis is based on a set of particle-mesh simulations,
carried out with MC$^3$ ({\bf M}esh-based {\bf C}osmology {\bf C}ode
on the {\bf C}ell), a new hybrid cosmology code that takes advantage
of Cell-accelerated hardware \citep{habib09,pope09}. We investigate a
$\Lambda$CDM model with the following cosmological parameters:
$\omega_m=0.1296$, $\omega_b=0.0224$, $n_S=0.97$, $\sigma_8=0.8$, and
$h=0.72$. We generate an initial condition with 256$^3$ particles on a
256$^3$ grid at a starting redshift $z=200$. We evolve these initial
conditions with two different uniform force grids: a 256$^3$ grid and
a 1024$^3$ grid. Therefore, the low resolution simulation has a force
resolution of $\sim$1$h^{-1}$Mpc and the higher resolution simulation,
of $\sim$250$h^{-1}$kpc. (Note that because this is a demonstration
problem, the chosen parameters are not representative of those
actually used in full-scale simulations.) For each simulation we store
the final timestep in the cosmo format which provides the positions
and velocities of the particles. For the mass field we store the
potential of each particle. These outputs can be readily read into
ParaView and be analyzed.

\subsection{The Analysis}

\begin{figure*}[t] \centerline{
    \includegraphics[width=3.5in]{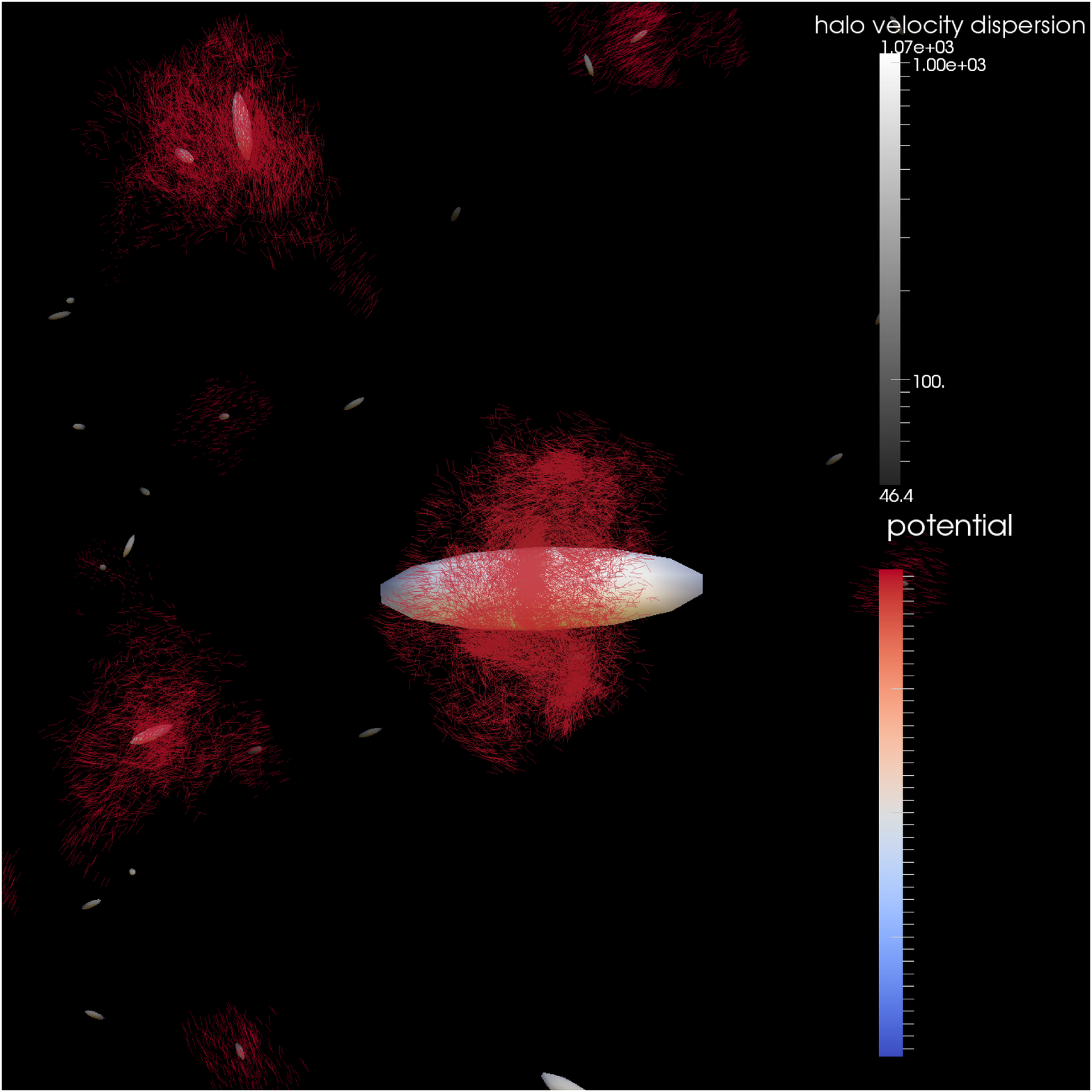}
    \includegraphics[width=3.5in]{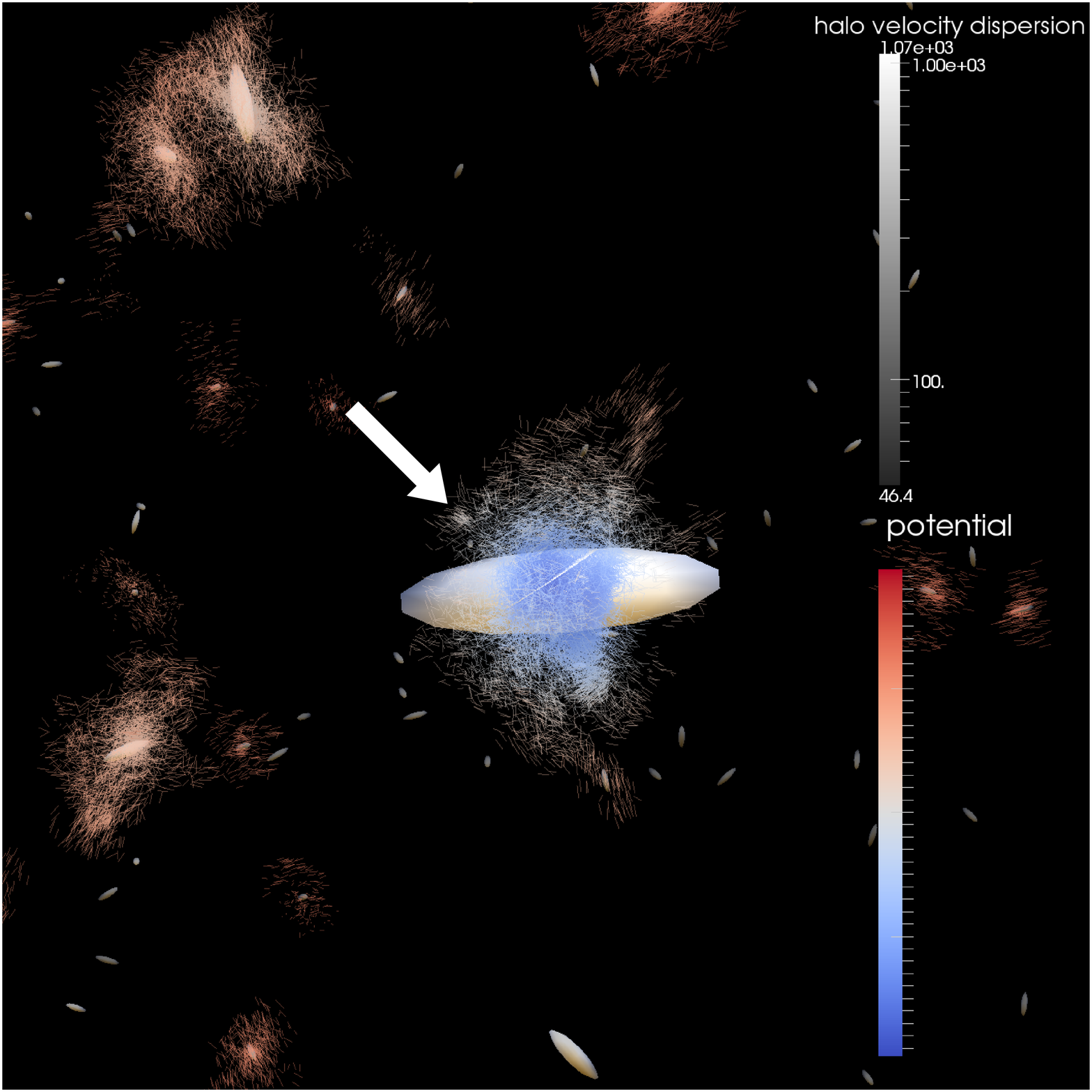}}
    \vspace{0.3cm} \caption{\label{halo1} Visualization of one of the
    largest halos in the simulations: lower resolution ($256^3$ grid,
    left), higher resolution ($1024^3$ grid, right). Shown are the
    halos themselves (ellipsoids, oriented with respect to velocity
    and colored with respect to velocity dispersion measured in km/s)
    and particles within halos (within a 20~$h^{-1}$Mpc box centered
    around the major halo). The particles are represented by glyphs
    pointing in the velocity direction and colored with respect to
    their potential value (arbitrary units). Halos which do not show
    particles around them are outside the chosen 20~$h^{-1}$Mpc
    box. The white arrow in the right plot points at a small subhalo
    missing in the lower resolution simulation plot on the left.}
\end{figure*}

In \cite{heitmann06} we derived a simple criterion for the force
resolution required to resolve the mass and position of halos with a
linking length $b=0.2$ reliably. This criterion states that
\begin{equation}\label{crit}
\frac{\delta_f}{\Delta_p} < 0.62\left[\frac{n_h\Omega(z)}{\Delta}\right]^{1/3},
\end{equation}
where $\delta_f$ is the force resolution (for a PM code,
$\delta_f/\Delta_p=n_p/n_g$ with $n_p$ being the cube root of the
number of particles and $n_g$ the cube root of the grid size), $n_h$
the number of particles per halo, $\Delta$ the overdensity of interest
measured with respect to the critical density, and $\Delta_p$ the
particle separation. A nominal choice of $\Delta=200$ corresponds
roughly to a ``virial mass''. With this choice, the criterion predicts
that the 256$^3$ grid simulation should have sufficient resolution to
capture halos with more than 3000 particles for $\Omega_m=0.25$. This
claim can be easily investigated with ParaView, and Figure~\ref{b02}
shows the results. We identify all halos with more than 3000 particles
(linking length of $b=0.2$) and show the particles that reside in
halos for the different resolutions in a sub-volume. In addition, as a
quantitative result, ParaView provides the overall number of halos
found (116 for the high resolution run, 110 for the low resolution
run) and we show a histogram of halo counts versus mass. This
histogram indicates that the 5\% discrepancy for the halo count is
dominated by the smaller halos. The very convenient feature now is
that once the analysis plot is set up as, e.g., shown in
Figure~\ref{b02}, we can readily change the parameters for the halo
finder and all panels will automatically show the new results. This
makes the analysis task fast and convenient, allowing exploration of
different settings in a simple manner. As an illustration, by changing
the linking length to $b=0.15$ in Figure~\ref{b015}, we find that the
high resolution simulation now has 84 halos compared to 65 halos in
the low resolution run, the difference between the two having
increased to 20\%. This is to be expected as it corresponds to an
effective increase of $\Delta$ in the denominator of
Eqn.~(\ref{crit}), making the inequality harder to satisfy. If we
lower the mass cut for the halos to 1000 particles per halo, we find,
for $b=0.15$, 487 halos in the high resolution run and 338 for the low
resolution run, a difference of 30\%.

Figure~\ref{massf} (also created within ParaView) summarizes the
results for halos with less than 2500 particles and a linking length
of $b=0.2$.  The red line represents the halo counts as function of
mass for the high resolution run, the black line the low resolution
run. Over the entire mass range, the high resolution run has more
halos.

As a next step we focus on a particular halo and investigate its
structure as a function of force resolution. We choose the heaviest
halo in the simulation that has no dominant substructure. The
visualization of this halo and the quantitative information available
from ParaView allows us to investigate the force resolution effects in
more detail. First, we compare the basic properties of the ($b=0.2$)
halo from the two simulations: Table~\ref{tab1} summarizes the
currently available halo properties as measured by ParaView. The halo
from the low resolution simulation is slightly heavier, though the
difference is below the percent level. The center of mass for both
halos is also very close, and differences are again below 1\%. The
force resolution effects become more apparent for the velocity
properties of the halos.  The center of mass velocities differs at the
10\% level and the high resolution halo has a larger velocity
dispersion, by about 20\%.

Next, we visualize the chosen halo and its surrounding region, as
shown in Figure~\ref{halo1}. To do this, we first identify all halos
in the simulations with more than 100 particles. We then focus on the
halo of interest and select all the particles that reside in halos in
a 20~$h^{-1}$Mpc box around the central halo. We zoom into the box
while continuing to display halos which reside behind it. The
particles are displayed as two-dimensional glyphs pointing in the
velocity direction of the particle. In addition, they are colored with
respect to their potential value -- red corresponds to a shallow
potential while blue corresponds to a deep potential. The color coding
is the same in both figures for ease of comparison. In addition to the
particles within the halos, we show the center of mass of each halo by
an ellipsoid pointing in the direction of the center of mass velocity.
The halos are colored by their measured velocity dispersion $\sigma_v$
(lighter colors correspond to higher values for $\sigma_v$) and sized
with respect to their mass. We therefore have the following
information about the halos depicted in this visualization: (i) the
number of halos in a certain region; (ii) the masses of halos; (iii)
their center of mass position and velocity; (iv) the halo velocity
dispersions; (v) velocity and position information about particles
within halos; (vi) the potential values of the halo particles.

\begin{table}\caption{Basic halo properties}
\begin{tabular}{l c c}
\hline
\hfill & Low resolution halo & High resolution halo \\
\hline
Mass [10$^{15}h^{-1}$M$_\odot$]&1.34406 & 1.34344\\
Center of Mass [$h^{-1}$Mpc] & (128.8, 85.5, 219.8) & (129.0, 85.6, 220.0)  \\
CoM velocity [km/s] & (-218.7, -94.2, -369.5) & (-191.3, -82.1, -367.3)\\
$\sigma_v$ [km/s] & 795.06 & 1072.16 \\
\hline \hline
\end{tabular}\label{tab1}
\end{table}

The first, and obvious, result of the local comparison is that the
lower resolution simulation has fewer halos overall. The next
immediate observation is the much deeper potential well at the center
of the high resolution central halo, as well as the deeper potentials
in the small neighboring halos. This deeper potential will lead to a
higher mass concentration in the center of the central halo. It is
also clear that the higher resolution simulation shows more
substructure, e.g., in the left upper part of the central halo the
high resolution result shows the formation of a small subhalo (marked
by the white arrow) which is absent in the low resolution
run. Overall, there are more particles on the outskirts of the low
resolution central halo; on the left side many more particles appear
to stream in to the halo. Thus, at least for this halo -- and
consistent with the overall results -- one may conclude that for
massive halos, the two force resolution effects compensate each other
and the halo mass remains robust. This result is also in good
agreement with the findings of \cite{heitmann05,CodeCompare} and
\cite{lukic}. In those papers, the mass functions obtained with
different codes were compared and good agreement established even
though the force resolution in the codes differed by up to a factor of
10. In \cite{bhattacharya} a more detailed study was carried out
analyzing halo mass differences from different force resolution
simulations. In that study, the difference in force resolution was
much larger than here -- two simulations with force resolutions
different by a factor of 14 were compared and the effect on the high
mass halos was found to be at 4\%. Overall, these findings are
encouraging with respect to obtaining accurate predictions for the
cluster mass function from moderate resolution simulations. This
relaxation of the spatial dynamic range requirement is particularly
useful for cluster simulations where a large volume is needed to get
good statistics for the associated mass function.

\section{Conclusion}
\label{conclusion}
In this paper we have introduced ParaView as a powerful and convenient
visualization and analysis tool for large cosmological $N$-body
simulations. ParaView is an open-source, parallel visualization
platform that can carry out visualization and analysis tasks on
desktops as well as on supercomputers. We have implemented new
readere and filters into ParaView that are designed for easy and
efficient analysis of cosmological simulations. These tools include
parallel particle readers (cosmo and {\small {\small GADGET}} format
are supported) and a very efficient halo finder. The underlying
infrastructure for the cosmology filters is taken from our recent code
developments for MC$^3$. As the analysis code suite for MC$^3$ evolves
and matures, we will port the new developments to ParaView. Currently,
a spherical overdensity halo finder and a sub-halo finder are under
final development.

In this paper, we demonstrated the use of ParaView and its interface
for analyzing and visualizing cosmological simulation with a few
examples, focusing on the effects of force resolution on the halo mass
function in the cluster regime. The strength of ParaView is the
ability of summarizing a large number of attributes of the simulation
in a compelling visualization and at the same time, allow for
visualization-aided analysis -- the availability of quantitative
information, allied to the visualization itself. Together with
manipulation and analysis tools such as a calculator and binning
routines, we believe that ParaView will be a very valuable new tool
for the cosmology community.

\acknowledgements

A special acknowledgment is due to supercomputing time awarded to us
under the LANL Institutional Computing Initiative. Part of this
research was supported by the DOE under contract W-7405-ENG-36. The
authors acknowledge support from the LDRD program at Los Alamos
National Laboratory.

\appendix

\begin{figure*}[t]
\centerline{
\includegraphics[width=6.7in]{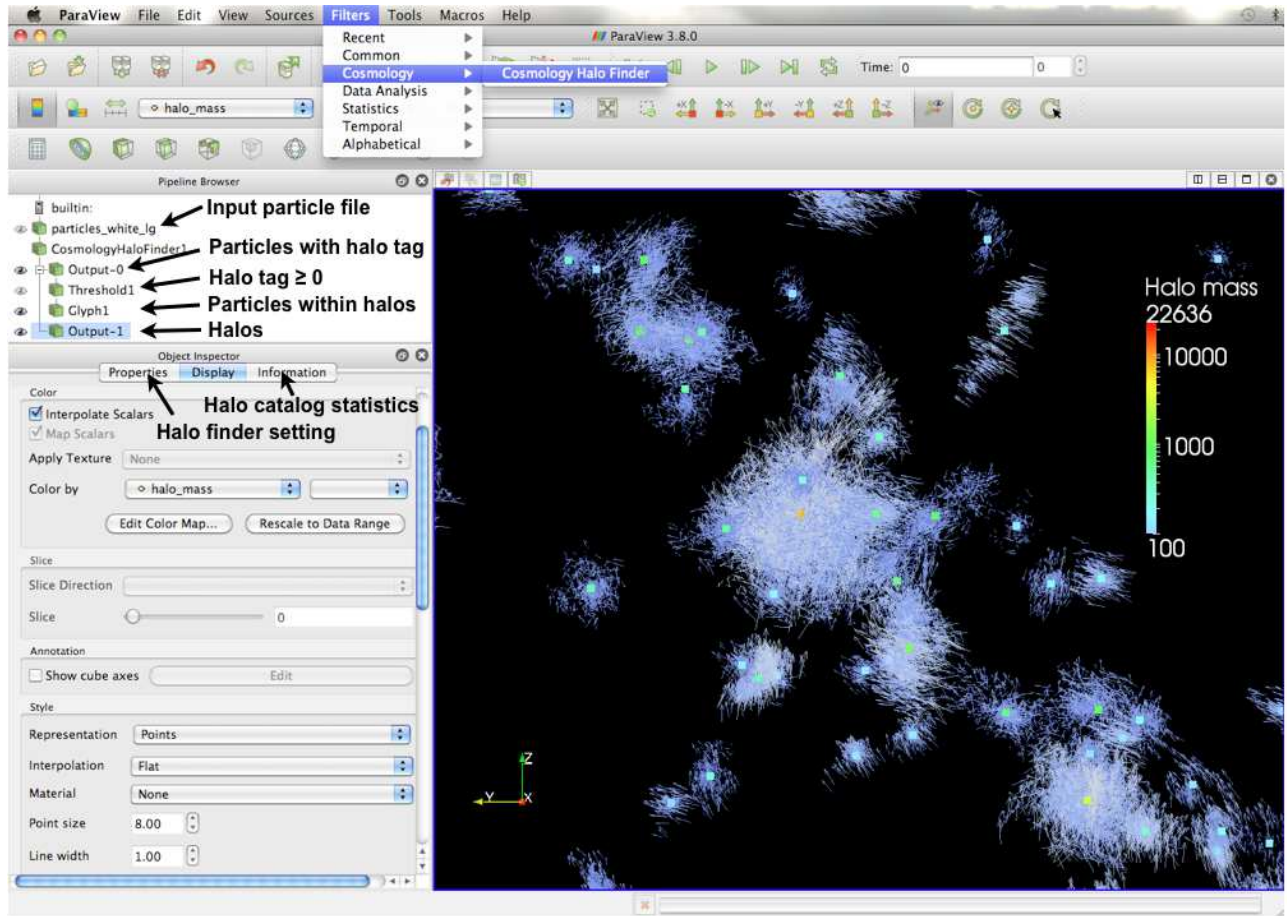}}
\vspace{0.3cm}
\caption{\label{screenshot_labeled}Screenshot of the ParaView
  GUI. Shown are particles within halos as two-dimensional glyphs and
  halo centers colored with respect to their mass.  ParaView allows
  for easy changes in the properties of the displayed particles (in
  this case, e.g., linking length and minimum number of particles in a
  halo for the halo finder), properties of the visualization itself,
  such as color schemes, and quantitative information about the data
  set, e.g., maximum and minimum positions, velocities, and tags of
  particles or halo counts.}
\end{figure*}

In this appendix we provide some usage tips on visualizing
cosmological $N$-body simulations with ParaView.
Figure~\ref{screenshot_labeled} shows a screenshot of the ParaView GUI
interface. The first step is to read the particle file of interest. If the
filename ends in .cosmo the ParaView reader will automatically
identify the file as cosmo format and choose the correct reader. If
the ending is different, a menu will appear and the user can pick by
hand which format the reader should use. Once the cosmo format is
specified, the user needs to enter the box size, the number of
particles in one dimension, and the overload length that should be
used for the halo finder. In the screenshot, the particle file that
was read in was called "particle\_white\_lg" as apparent from the left
upper list. Once the file is read in, some information is readily
available and can be accessed by using the Information tab (e.g.
number of particles, minimum and maximum velocities and positions).
The Display tab opens choices for displaying the particles -- the
default option under ``Style'' and then ``Representation'' is
``Outline'' which will simply draw a box around the whole particle
distributions. Changing this option to ``Points'' (as is done in the
figure) will display the actual particles. Some of the particle
attributes can then be changed, e.g. the size and the color options.
The ``eye'' next to the particles\_white\_lg can be activated or
de-activated by clicking on it -- in our example it is de-activated
which means that we do not show all the particles from the simulation,
as explained below.

Next, we can apply a filter to the data as shown in the upper part of
the figure. In the example, we evoke the halo finder. As in the case
of the particle reader, options appear under ``Properties" so that the
linking length, minimum number of particles per halo and overload
length can be specified. The halo finder then generates two new output
files, ``Output-0'' and ``Output-1''. The first file holds all
particles with the additional information of the halo tag. Particles
which are not in a halo have the tag -1. By using another filter,
``threshold'' and requesting only particles to be displayed with halo
tags $\ge$0, all particles in halos can be displayed. In the example,
we decided to show these particles as two-dimensional arrows colored
with respect to velocities. In order to do this, we used another
filter ``Glyphs'' which allows for this option. As the activated eye
next to ``Glyph1'' shows we are displaying these glyphs in the main
window. ``Output-1'' contains the actual halo catalog. Again, by
choosing ``Information'', measurements of halo properties such as mass
ranges and velocities will be shown. In the figure, ``Output-1'' is
shaded in blue, which means the menu below can be manipulated for that
output. In the current case, the halos are shown as points of size 8
colored with respect to mass.


\begin{thebibliography}{99.}

\bibitem[{{Ahrens et al.}(2005)}]{paraview}Ahrens,~J., Geveci,~B., \&
  Law,~C. 2005, ParaView: An End-User Tool for Large Data Visualization. In
  the Visualization Handbook. Edited by Hansen,~C.D. and
  Johnson,~C.R. Elsevier.

\bibitem[{{Bhattacharya et al.}(2010)}]{bhattacharya} Bhattacharya,~S.,
  Heitmann,~K., White,~M., Wagner,~C., Luki\'c,~Z., \& Habib,~S. 2010,
  arXiv:1005.2239 [astro-ph.CO]

\bibitem[{{Hsu et al.}(2010)}]{Vis2010} Hsu,~C.-H., Ahrens,~J., \&
  Heitmann,~K. 2010, Pacific Visualization 2010

\bibitem[{{Crocce et al.}(2010)}]{MICE} Crocce,~M., Fosalba,~P.,
  Castander,~F.J., Gaztanaga,~E. 2010, MNRAS, 403, 1353

\bibitem[{{Cunha \& Evrard}(2010)}]{cunha} Cunha,~C. \& Evrard,~A.
2010, Phys. Rev. D81, 083509

\bibitem[{{Habib et al.}(2009)}]{habib09} Habib,~S. et al. 2009,
  Journal of Physics: Conference Series, 180, 012019

\bibitem[{{Heitmann et al.}(2006)}]{heitmann06} Heitmann,~K.,
  Luki\'c,~Z., Habib,~S., \& Ricker,~P. 2006, ApJ, 642, L85

\bibitem[{{Heitmann et al.}(2005)}]{heitmann05} Heitmann,~K.,
  Ricker,~P.M., Warren,~M.S., \& Habib,~S. 2005, ApJS, 160, 28

\bibitem[{{Heitmann et al.}(2008)}]{CodeCompare} Heitmann,~K. et
  al. 2008, Computational Science and Discovery, 1, 015003

\bibitem[{{Heitmann et al.}(2010)}]{CoyoteI} Heitmann,~K., White,~M.,
  Wagner,~C., Habib,~S., \& Higdon,~D. 2010, ApJ, 715, 104

\bibitem[{{Heitmann et al.}(2009)}]{CoyoteII} Heitmann,~K.,
  Higdon,~D., White,~M., Habib,~S., Williams,~B., \& Wagner,~C. 2009,
  ApJ, 705, 156

\bibitem[{{Kim et al.}(2009)}]{horizon2} Kim,~J., Park,~C., Gott,
  III,~R., Dubinski,~J. 2008,  ApJ, 701, 1547

\bibitem[{{Lawrence et al.}(2010)}]{CoyoteIII} Lawrence,~E.,
  Heitmann,~K., White,~M., Higdon,~D., Wagner,~C., Habib,~S., \&
  Williams,~B. 2010, ApJ, 713, 1322

\bibitem[{{Luki\'c et al.}(2007)}]{lukic} Luki\'c, Z., Reed, D.,
  Habib, S., \& Heitmann, K. 2009, ApJ 692, 217

\bibitem[{{McBride et al.}(2010)}]{mcbride2010}McBride, C. et al.
  2010, in preparation; see also 
http://lss.phy.vanderbilt.edu/lasdamas/overview.html

\bibitem[{{Navarro et al.}(1997)}]{NFW} Navarro, J., Frenk, C.S., \&
  White, S.D.M. 1997, ApJ 490, 493

\bibitem[{{Pope et al.}(2010)}]{pope09} Pope,~A., Habib,~S.,
  Lukic,~Z., Daniel,~D., Fasel,~P., Desai,~N. \& Heitmann,~K. 2010,
  Computing in Science and Engineering 12, 17

\bibitem[{{Schroeder et al.}(1996)}]{vtk} Schroeder,~W.J.,
  Martin,~K.M., \& Lorensen,~W.E., 1996, The Visualization Toolkit: An
  Object Oriented Approach to 3D Graphics, Prentice Hall

\bibitem[{{Springel}(2005)}]{springel} Springel,~V. 2005, MNRAS, 364,
  1105
	
\bibitem[{{Summerfield}(2010)}]{Qt} Summerfield,~M. 2010, Advanced Qt
  Programming: Creating Great Software with C++ and Qt, Addison-Wesley
	
\bibitem[{{Teyssier et al.}(2009)}]{horizon1} Teyssier,~R., Pires,~S.,
  Prunet,~S., Aubert,~D., Pichon,~C., Amara,~A., Benabed,~K.,
  Colombi,~S., Refregier,~A., \& Starck,~J.-L. 2009, A\&A, 497, 335

\bibitem[{{Warren et al.}(2006)}]{warren} Warren,~M.S., Abazaijan,~K.,
  Holz,~D.E., \& Teodoro,~L. 2006, ApJ 646, 881

\bibitem[{{Wu et al.}(2010)}]{heidi} Wu,~H.-Y., Zentner,~A.R., \&
  Wechsler,~R.H. 2010, ApJ, 713, 856


\end{thebibliography}
\end{document}